\documentclass[preprintnumbers,twocolumn,prl,aps,superscriptaddress,footinbib,amsfonts,amsmath,amssymb,bm,showpacs,floatfix]{revtex4-1}

\usepackage[T1]{fontenc}
\usepackage{blindtext}
\usepackage{amsmath}
\usepackage{amssymb}
\usepackage{graphics,graphicx}
\usepackage{dcolumn}
\usepackage{bm}
\usepackage{epsfig}
\usepackage[normalem]{ulem}
\usepackage[usenames]{color}
\usepackage{mathbbol}
\usepackage{epstopdf}
\usepackage{simplewick}
\usepackage[utf8]{inputenc} 
\usepackage{slashed}
\usepackage{soul}
\usepackage[dvipsnames]{xcolor}
\usepackage[export]{adjustbox}
\usepackage{mathtools}

\allowdisplaybreaks

\begin{document}

\preprint{JLAB-THY-23-3826}

\title{\bf Number density interpretation of dihadron fragmentation functions}

\newcommand*{\LVC}{Department of Physics, Lebanon Valley College, Annville, Pennsylvania 17003, USA}\affiliation{\LVC}
\newcommand*{\TU}{Department of Physics, SERC, Temple University, Philadelphia, Pennsylvania 19122, USA}\affiliation{\TU}
\newcommand*{\PSU}{Division of Science, Penn State University Berks, Reading, Pennsylvania 19610, USA}\affiliation{\PSU}
\newcommand*{\JLAB}{Jefferson Lab, Newport News, Virginia 23606, USA}\affiliation{\JLAB}

\author{D.~Pitonyak}\affiliation{\LVC}
\author{C.~Cocuzza}\affiliation{\TU}
\author{A.~Metz}\affiliation{\TU}
\author{A.~Prokudin}\affiliation{\PSU}\affiliation{\JLAB}
\author{N.~Sato}\affiliation{\JLAB}

\begin{abstract}
\noindent We present a new quantum field-theoretic definition of fully unintegrated dihadron fragmentation functions (DiFFs) as well as a generalized version for $n$-hadron fragmentation functions.   
We demonstrate that this definition allows certain sum rules to be satisfied, making it consistent with a number density interpretation.  Moreover, we show how our corresponding so-called extended DiFFs  that enter existing phenomenological studies are number densities and also derive their evolution equations.  Within this new framework, DiFFs extracted from experimental measurements will have a clear physical meaning.
\end{abstract}

\maketitle 

{\it Introduction ---}  High-energy collisions of hadrons are central to understanding their femtoscale structure at the level of quarks and gluons (partons) within the theory of quantum chromodynamics (QCD).  The critical ingredients that encode this information are parton distribution functions (PDFs) and fragmentation functions (FFs). A crucial property of PDFs and FFs is their interpretation as number densities in a parton model framework~\cite{Collins:1981uw,Collins:2011zzd}, which consequently allows one to  derive certain sum rules~\cite{Collins:1981uw,Collins:2011zzd,Meissner:2010cc,Accardi:2020iqn,Ito:2009zc}. For example, the unpolarized transverse momentum dependent (TMD) PDF $f_1^{i/N}(x,\vec{k}_T^2)$ gives the number density in the momentum fraction $x$ and transverse momentum $\vec{k}_T$ of a  parton $i = q\;{\rm or}\; g$ in a  nucleon~$N$~\cite{Collins:1981uw,Collins:2011zzd}.  Similarly, the unpolarized TMD FF $D^{h/i}_1(z,\vec{P}_\perp^2)$ gives the number density in the momentum fraction $z$ and transverse momentum $\vec{P}_\perp$ of a  hadron $h$ fragmenting from a  parton $i$~\cite{Collins:1981uw,Collins:2011zzd}. Since PDFs and FFs are number densities,  one can also use them to calculate expectation values (see, e.g., Refs.~\cite{Sato:2019yez,Hou:2019efy,Bacchetta:2022awv,Barry:2023qqh}).  The information contained in sum rules and expectation values are  important pieces to understanding hadronic structure as well as constraining or cross-checking phenomenological extractions and model calculations of PDFs and FFs. 

The most common type of FFs describes the situation where a single hadron $h$ is detected in the final state, $i\to hX$ ($X$ representing all undetected particles).  Another intensely studied class of reactions analyzes the case of two hadrons $h_1, h_2$ being detected from the same parton-initiated jet, $i\to (h_1h_2)X$, where dihadron FFs (DiFFs) become relevant~\cite{Konishi:1978yx,Konishi:1979cb,Sukhatme:1980vs,Vendramin:1980wz,Vendramin:1981te,Sukhatme:1981ym,Collins:1993kq,Collins:1994ax,Jaffe:1997hf,Jaffe:1997pv,Bianconi:1999cd,Bianconi:1999uc,Radici:2001na,Bacchetta:2002ux,Bacchetta:2003vn,Boer:2003ya,deFlorian:2003cg,Bacchetta:2004it,Majumder:2004wh,Majumder:2004br,Ceccopieri:2007ip,Bacchetta:2008wb,Bacchetta:2011ip,Casey:2012ux,Casey:2012hg,Courtoy:2012ry,Courtoy:2012ry,Bacchetta:2012ty,Matevosyan:2013aka,Gliske:2014wba,Radici:2015mwa,Metz:2016swz,Radici:2016lam,Matevosyan:2017alv,Matevosyan:2017uls,Matevosyan:2017liq,Radici:2018iag,Matevosyan:2018jht,Matevosyan:2018icf,Benel:2019mcq,Chen:2022pdu,Chen:2022muj}.  
The quantum field-theoretic definition of DiFFs at the fully unintegrated level (what we will call uDiFFs) was first written down 
almost 25 years ago 
in the pioneering paper of Bianconi, Boffi, Jakob, and Radici (BBJR)~\cite{Bianconi:1999cd}.  This work has been the basis for all subsequent dihadron-related research for observables sensitive to the relative transverse momentum of the two hadrons~\cite{Bianconi:1999uc,Radici:2001na,Bacchetta:2002ux,Bacchetta:2003vn,Boer:2003ya,Bacchetta:2004it,Bacchetta:2008wb,Bacchetta:2011ip,Courtoy:2012ry,Courtoy:2012ry,Bacchetta:2012ty,Matevosyan:2013aka,Gliske:2014wba,Radici:2015mwa, Radici:2016lam,Matevosyan:2017alv,Matevosyan:2017uls,Matevosyan:2017liq,Radici:2018iag,Matevosyan:2018jht,Matevosyan:2018icf,Benel:2019mcq}.  
Unfortunately, the BBJR definition  does not allow the uDiFFs, nor the so-called extended DiFFs (extDiFFs) that are the focus of existing phenomenological analyses, 
to retain a number density interpretation in a parton model framework.

The main purpose of this Letter is to disseminate a new definition of uDiFFs that corrects this issue. We justify its number density interpretation by explicitly proving certain sum rules. 
We also show our corresponding extDiFFs are number densities and derive their evolution equations.   Given the existing electron-positron annihilation dihadron cross section data~\cite{Belle:2017rwm}, dihadron transverse single-spin asymmetries in electron-positron annihilation~\cite{Belle:2011cur}, semi-inclusive deep-inelastic scattering~\cite{HERMES:2008mcr,COMPASS:2023cgk}, and proton-proton collisions~\cite{STAR:2015jkc,STAR:2017wsi}, and anticipated measurements of the proton-proton dihadron cross section and SIDIS dihadron multiplicities, one eventually will be able to perform rigorous fits of extDiFFs within QCD global analyses. These studies must be carried out within our new framework for the extracted extDiFFs to have a clear physical meaning -- see Refs.~\cite{Cocuzza:2023oam,Cocuzza:2023vqs}.

{\it New Correlator Definition of DiFFs---} We begin by briefly discussing two different reference frames that will be relevant for our analysis:~the ``parton frame'' (p), where the fragmenting parton has no transverse momentum, and the ``dihadron frame'' (h), where the dihadron has no transverse momentum.  In both frames the parton has the same large minus-lightcone momentum component $k^-$ ($V^\pm \equiv (V^0\pm V^3)/\sqrt{2}$ for a generic vector $V$).  They are connected through the following Lorentz transformation~(see, e.g., Ref.~\cite{Collins:2011zzd} Sec.~12.4.1):~$V_{\rm p}^-=V_{\rm h}^-\equiv V^-$; $V_{\rm p}^+ = (\vec{k}_T/k^-)^2\, V^-/2+V_{\rm h}^+-\vec{k}_T\cdot \vec{V}_T/k^-$; $\vec{V}_\perp = -(\vec{k}_T/k^-)V^-+\vec{V}_T$. We use $\perp (T)$ to denote transverse components in the parton (dihadron) frame.
The parton frame is more natural for the formulation of fragmentation correlators (whether single hadron or dihadron) as number densities, whereas the dihadron frame is more practical for proofs of factorization needed for phenomenological applications.

The quantum field-theoretic correlator for the fragmentation of a parton $i$ into two hadrons $h_1,h_2$, after integrating over $k^+$,
is defined as~\cite{Bianconi:1999cd}
\begin{align}
    \Delta_{\alpha\beta}^{h_1h_2/i}&(z_1,z_2,\vec{P}_{1\perp},\vec{P}_{2\perp})  \label{e:DiFF_corr} \\  
    &=\frac{1}{N_i}\!\sum_X\hspace{-0.5cm}\int\! \int\!\frac{d\xi^+d^2\vec{\xi}_\perp}{(2\pi)^3}\,e^{ik\cdot\xi}\,\mathcal{O}^{h_1h_2/i}_{\alpha\beta}(\xi)\Big |_{\xi^-=0}\,,\nonumber 
\end{align}
where $z_1$, $z_2$ are the fractions of the parton's longitudinal momentum  carried by each hadron, and $\vec{P}_{1\perp},\vec{P}_{2\perp}$ are the  transverse momenta of the hadrons 
relative to the parton.
For a quark, $N_i$ is the number of quark colors $N_c=3$, and
\begin{align}
    \mathcal{O}^{h_1h_2/q}_{\alpha\beta}(\xi)&= \langle 0|\mathcal{W}(\infty,\xi)\psi_{q,\alpha}(\xi^+,0^-,\vec{\xi}_\perp)|P_1,P_2;X\rangle \nonumber\\
   &\hspace{-1cm}\times \langle P_1,P_2;X|\bar{\psi}_{q,\beta}(0^+,0^-,\vec{0}_\perp)\mathcal{W}(0,\infty)|0\rangle\,,\label{e:qop}
\end{align}
where $\psi_q$ is the quark field, $\alpha,\beta$ are  indices for the components of the field, and $\mathcal{W}$ is a  Wilson line in the fundamental representation of SU(3) that ensures color gauge invariance~\cite{Boer:2003cm,Collins:2011zzd}.  A sum over color indices in Eq.~(\ref{e:qop}) is implied.
For a gluon, $N_i=N_c^2-1$, and
\begin{align}
   \mathcal{O}^{h_1h_2/g}_{\alpha\beta}(\xi) &= \langle 0|\mathcal{W}^{ba}(\infty,\xi)F^{a}_{+\alpha}(\xi^+,0^-,\vec{\xi}_\perp)|P_1,P_2;X\rangle \nonumber\\
   &\hspace{-1cm}\times\, 
   \langle P_1,P_2;X|F^c_{+\beta}(0^+,0^-,\vec{0}_\perp)\mathcal{W}^{cb}(0,\infty)|0\rangle\,,\label{e:gop}
\end{align}
where 
$F^a_{\mu\nu}=\partial_\mu A_\nu^a-\partial_\nu A_\mu^a+gf^{abc}A_\mu^b A_\nu^c$ is the field strength tensor involving the gluon field $A$, and the Wilson lines are now in the adjoint representation of SU(3).

Throughout this Letter we focus on the production of unpolarized hadrons.  For the fragmentation of an unpolarized parton, we parameterize the correlator in Eq.~(\ref{e:DiFF_corr}) as
\begin{align}
    &\frac{1}{64\pi^3z_1 z_2}{\rm Tr}\!\left[\Delta^{h_1h_2/q}(z_1,z_2,\vec{P}_{1\perp},\vec{P}_{2\perp})\gamma^-\right]\label{e:D1q_uDiFF}\\
    &\hspace{2cm}=D_1^{h_1h_2/q}(z_1,z_2,\vec{P}_{1\perp}^2,\vec{P}_{2\perp}^2,\vec{P}_{1\perp}\!\cdot \!\vec{P}_{2\perp})\,,\nonumber\\[0.1cm]
    &\frac{z}{32\pi^3z_1z_2P_h^-}\,\delta_\perp^{ij} \,\Delta^{h_1h_2/g,ij}(z_1,z_2,\vec{P}_{1\perp},\vec{P}_{2\perp}) \label{e:D1g_uDiFF}\\
    &\hspace{2cm}= D_1^{h_1h_2/g}(z_1,z_2,\vec{P}_{1\perp}^2,\vec{P}_{2\perp}^2,\vec{P}_{1\perp}\!\cdot\! \vec{P}_{2\perp})\,,\nonumber
\end{align}
where $z= z_1+z_2$ is the total momentum fraction of the dihadron and $P_h = P_1+P_2$.  
As we will show in the next section, the prefactor of $1/(64\pi^3z_1 z_2)$ in Eq.~(\ref{e:D1q_uDiFF}) is crucial to justifying the number density interpretation of the quark uDiFFs (and similarly for the gluon case in Eq.~(\ref{e:D1g_uDiFF})).  If one instead were to use a prefactor of $1/(4z)$, to be in full analogy with single-hadron fragmentation~\cite{Collins:1981uw,Mulders:1995dh,Boer:2003cm,Bacchetta:2006tn}, the quark uDiFFs would not retain a number density interpretation.  Indeed, the fact that the prefactors on the l.h.s.~of Eqs.~(\ref{e:D1q_uDiFF}), (\ref{e:D1g_uDiFF}) are needed was already recognized previously in the context of collinear DiFFs $D_1^{h_1h_2/i}(z_1,z_2)$~\cite{Majumder:2004wh,Majumder:2004br}.  

{\it Number Density Interpretation ---} 
To justify that Eqs.~(\ref{e:D1q_uDiFF}), (\ref{e:D1g_uDiFF}) have the desired number density interpretation, we will derive sum rules involving our uDiFFs in a parton model framework. The proofs of the sum rules in this section are left for Supplemental Material. We focus first on the number sum rule, 
\begin{align}
    \int \!\!d\mathcal{PS}\,D_1^{h_1h_2/i}(z_1,z_2,\vec{P}_{1\perp}^{\,2},\vec{P}_{2\perp}^{\,2},\vec{P}_{1\perp}\!\!\cdot\! \vec{P}_{2\perp})
    = \langle \mathcal{N}(\mathcal{N}-1)\rangle\,,\label{e:num_sumrule}
\end{align}
where $\int\! d\mathcal{PS}=\sum_{h_1}\sum_{h_2}\! \int_0^1 \!dz_2 \int_0^{1-z_2} \!\!dz_1 \!\int \!\!d^2\vec{P}_{1\perp}\int \!\!d^2\vec{P}_{2\perp}$, and $\mathcal{N}$ is the total number of hadrons produced when the parton $i$ fragments. Thus,  $\langle\mathcal{N}(\mathcal{N}-1)\rangle$ is the expectation value for the total number of hadron pairs produced in the fragmentation of $i$. A sum over hadron spins must be included if either or both hadrons have nonzero spin.  We remark that the labeling of the two hadrons as $(h_1,h_2)$ or $(h_2,h_1)$ is distinguishable and no factor of $1/2$ is needed in the r.h.s.~of Eq.~(\ref{e:num_sumrule}). We note that the number sum rule Eq.~(\ref{e:num_sumrule}) was first derived in Ref.~\cite{Majumder:2004br}.  A crucial step in our proof is being able to introduce the number operator, 
\begin{equation}
    \hat{N}_{h_j} \equiv \int\!\frac{dP_j^-d^2\vec{P}_{j\perp}}{(2\pi)^3 \,2P_j^-}\,\hat{a}_{h_j}^\dagger \hat{a}_{h_j} = \int\!\frac{dz_jd^2\vec{P}_{j\perp}}{(2\pi)^3 \,2z_j}\,\hat{a}_{h_j}^\dagger \hat{a}_{h_j} \,,
\end{equation}
for each hadron ($j=1\;{\rm or}\;2$).  This can only be achieved by having the specific prefactors on the l.h.s.~of Eqs.~(\ref{e:D1q_uDiFF}), (\ref{e:D1g_uDiFF}). Indeed, a derivation is not possible if a prefactor of $1/(4z)=1/(4(z_1+z_2))$ is used on the l.h.s.~of Eq.~(\ref{e:D1q_uDiFF}).

The result in Eq.~(\ref{e:num_sumrule}) gives a clear interpretation for the uDiFFs we defined in Eqs.~(\ref{e:D1q_uDiFF}), (\ref{e:D1g_uDiFF}):~they are densities in the momentum fractions $z_1,z_2$ and transverse momenta $\vec{P}_{1\perp},\vec{P}_{2\perp}$ for the number of hadron pairs $(h_1 h_2)$ fragmenting from a parton $i$.  The uDiFF $D_1^{h_1h_2/q}(z_1,z_2,\vec{P}_{1\perp}^{\,2},\vec{P}_{2\perp}^{\,2},\vec{P}_{1\perp}\!\cdot\! \vec{P}_{2\perp})$ encodes the dihadron fragmentation process for an unpolarized quark ($\gamma^-$ projection of the correlator).  The number density interpretation also holds for the fragmentation of a longitudinally polarized quark ($\gamma^-\gamma^5$ projection) and a transversely polarized quark ($i\sigma^{i-}\gamma_5$ projection). The explicit parameterization of Eq.~(\ref{e:DiFF_corr}) in terms of quark and gluon uDiFFs for all parton polarizations, as functions of $(z_1,z_2,\vec{P}_{1\perp}^{\,2},\vec{P}_{2\perp}^{\,2},\vec{P}_{1\perp}\!\cdot\! \vec{P}_{2\perp})$, is given in Supplemental Material. 

We can also derive a momentum sum rule involving uDiFFs and TMD FFs, 
\begin{align}
    \sum_{h_1}\!\! \int_0^{1-z_2} \!\!\!dz_1 \!&\!\int \!\!d^2\vec{P}_{1\perp}\,z_1\,D_1^{h_1h_2/i}(z_1,z_2,\vec{P}_{1\perp}^{\,2},\vec{P}_{2\perp}^{\,2},\vec{P}_{1\perp}\!\!\cdot\! \vec{P}_{2\perp})\nonumber\\[-0.3cm]
    &= (1-z_2)\,D_1^{h_2/i}(z_2,\vec{P}_{2\perp}^{\,2})\,.\label{e:sumrule_unint}
\end{align}
If either or both hadrons have nonzero spin, then a sum over the spin of $h_1$ must be included on the l.h.s.~of Eq.~(\ref{e:sumrule_unint}) (and Eq.~(\ref{e:sumrule_cDiFF}) below). Note that one can identify the ratio of the uDiFF to the TMD FF, $D_1^{h_1h_2/i}(z_1,z_2,\vec{P}_{1\perp}^{\,2},\vec{P}_{2\perp}^{\,2},\vec{P}_{1\perp}\!\!\cdot\! \vec{P}_{2\perp})/D_1^{h_2/i}(z_2,\vec{P}_{2\perp}^{\,2})$, as a conditional number density in the momentum $(z_1,\vec{P}_{1\perp})$ for $h_1$ fragmenting from $i$ given $h_2$ has fragmented from $i$ with momentum $(z_2,\vec{P}_{2\perp})$.  
Further integrating Eq.~(\ref{e:sumrule_unint}) over $\vec{P}_{2\perp}$
yields
\begin{equation}
    \sum_{h_1}\!\! \int_0^{1-z_2} \!\!dz_1 \,z_1\,D_1^{h_1h_2/i}(z_1,z_2) = (1-z_2)D_1^{h_2/i}(z_2)\,.\label{e:sumrule_cDiFF}
\end{equation}
The momentum sum rule Eq.~(\ref{e:sumrule_cDiFF}) was first put forth in Refs.~\cite{Konishi:1979cb,Vendramin:1981te}. 
We also mention that the study of DiFFs has a close connection to double PDFs (DPDFs), where two partons emerge from a single nucleon.  Indeed, an analogous sum rule to Eq.~(\ref{e:sumrule_cDiFF}) exists for DPDFs, as was derived in Refs.~\cite{Gaunt:2009re,Diehl:2018kgr}. 
The quantum field-theoretic derivation of the sum rule Eq.~(\ref{e:sumrule_unint}) at the unintegrated (transverse momentum dependent) operator level (from which Eq.~(\ref{e:sumrule_cDiFF}) follows immediately) is a new aspect presented here for the first time.

One can readily generalize to $n$-hadron ($n\ge 1$) fragmentation in a way that retains a  number density interpretation:
\begin{align}
    &\frac{1}{4(16\pi^3)^{n-1}z_1\cdots z_n}{\rm Tr}\!\left[\Delta^{\{h_i\}_n/q}(\{z_i\}_n,\{\vec{P}_{i\perp}\}_n)\gamma^-\right]\nonumber\\
    &\hspace{1cm}= D_1^{\{h_i\}_n/q}(\{z_i\}_n,\{\vec{P}_{i\perp}^2\}_n,\{\vec{P}_{i\perp}\!\cdot\vec{P}_{j\perp}\}_n)\,,\label{e:uDiFF1p_n}\\[0.2cm]
    &\frac{z}{2P_h^-(16\pi^3)^{n-1}z_1\cdots z_n}\,\delta_\perp^{ij} \,\Delta^{\{h_i\}_n/g,ij}(\{z_i\}_n,\{\vec{P}_{i\perp}\}_n)\nonumber\\  &\hspace{1cm}= D_1^{\{h_i\}_n/g}(\{z_i\}_n,\{\vec{P}_{i\perp}^2\}_n,\{\vec{P}_{i\perp}\!\cdot\vec{P}_{j\perp}\}_n) \,,\label{e:D1g_uDiFF_n}
\end{align}
where $z=z_1+\cdots +z_n$, $P_h=P_1+\cdots +P_n$, $\{h_i\}_n\equiv h_1\cdots h_n$, $\{z_i\}_n\equiv z_1,\dots,z_n$, $\{\vec{P}_{i\perp}\}_n\equiv \vec{P}_{1\perp},\dots,\vec{P}_{n\perp}$, $\{\vec{P}_{i\perp}^2\}_n\equiv \vec{P}_{1\perp}^2,\dots,\vec{P}_{n\perp}^2$, $\{\vec{P}_{i\perp}\cdot\vec{P}_{j\perp}\}_n\equiv \vec{P}_{1\perp}\!\cdot \!\vec{P}_{2\perp},\dots,\vec{P}_{1\perp}\cdot \!\vec{P}_{n\perp},\vec{P}_{2\perp}\cdot \!\vec{P}_{3\perp},\dots,\vec{P}_{2\perp}\cdot \!\vec{P}_{n\perp},\,{\rm etc.}$  The correlators $\Delta^{\{h_i\}_n/i}(\{z_i\}_n,\{\vec{P}_{i\perp}\}_n)$ are the natural extensions of Eqs.~(\ref{e:qop}), (\ref{e:gop}) to $n$ hadrons, i.e., the final state is now $|P_1,\dots,P_n;X\rangle$.  The corresponding number sum rule reads
\begin{align}
   \!\int \!\!d\mathcal{PS}_n\, D_1^{\{h_i\}_n/i}(\{\cdots\}_n\!)= \!\left\langle\prod_{k=0}^{n-1}(\mathcal{N}-k)\!\right\rangle
    \,,\label{e:num_sumrule_n}
\end{align}  
where $\int \!\!d\mathcal{PS}_n$ denotes the $n$-hadron version of $\int \!\!d\mathcal{PS}$, 
and we have abbreviated the arguments of the FF.  Interestingly, the evolution of collinear $n$-hadron FFs was already studied some time ago~\cite{Vendramin:1980wz,Sukhatme:1980vs}, as well as more recently in Refs.~\cite{Chen:2022pdu,Chen:2022muj}, but no correlator definition was presented.

{\it Connection to Phenomenology} ---
In order to analyze measurements of dihadron observables, it becomes convenient to change to the dihadron frame~\cite{Bianconi:1999cd,Bacchetta:2002ux}. 
In addition to $P_h$, we also introduce the relative momentum $R=(P_1-P_2)/2$.  The individual hadrons have masses $M_1$ and $M_2$, while the invariant mass (squared) of the dihadron is $M_h^2=P_h^2$.  
Along with $z$, we form the variable $\zeta=(z_1-z_2)/z$.  The hadron momenta $P_1$ and $P_2$ can then be written as  $P_1 = \left(\frac{M_1^2+\vec{R}_T^{\,2}}{(1+\zeta)P_h^-},\frac{1+\zeta}{2}P_h^-,\vec{R}_T\right)$ and 
$P_2 = \left(\frac{M_2^2+\vec{R}_T^{\,2}}{(1-\zeta)P_h^-},\frac{1-\zeta}{2}P_h^-,-\vec{R}_T\right)$.
Note that one readily finds 
$\vec{R}_T^{\,2} = \frac{1-\zeta^2}{4}M_h^2 - \frac{1-\zeta}{2}M_1^2-\frac{1+\zeta}{2}M_2^2$. 
Due to this change of reference frames, one naturally thinks of uDiFFs as now depending on $(z,\zeta,\vec{k}_T^{\,2},\vec{R}_T^{\,2},\vec{k}_T\cdot \vec{R}_T)$ rather than $(z_1,z_2,\vec{P}_{1\perp}^{\,2},\vec{P}_{2\perp}^{\,2},\vec{P}_{1\perp}\!\!\cdot\! \vec{P}_{2\perp})$.

Nevertheless, the form of the number sum rule in Eq.~(\ref{e:num_sumrule}) allows us to generalize the idea of uDiFFs as number densities to any set of variables we choose.  Consider making a change of variables from $(z_1,z_2,\vec{P}_{1\perp},\vec{P}_{2\perp})$ to $(w,x,\vec{Y},\vec{Z})$, where we understand $w,x$ to be scalars and $\vec{Y},\vec{Z}$ to be two-dimensional vectors.  Then Eq.~(\ref{e:num_sumrule}) implies 
\begin{align}
D_1^{h_1 h_2/i}&(w,x,\vec{Y}^2,\vec{Z}^2,\vec{Y}\!\cdot\!\vec{Z})\nonumber \\
&\equiv \mathcal{J}\cdot D_1^{h_1 h_2/i}(z_1,z_2,\vec{P}_{1\perp}^2,\vec{P}_{2\perp}^2,\vec{P}_{1\perp}\!\cdot\! \vec{P}_{2\perp}) \label{e:JuDiFF}
\end{align}
is a number density in $(w,x,\vec{Y},\vec{Z})$, where $\mathcal{J}=|\partial(z_1,z_2,\vec{P}_{1\perp},\vec{P}_{2\perp})/\partial (w,x,\vec{Y},\vec{Z})|$ is the Jacobian for the change of variables from $(z_1,z_2,\vec{P}_{1\perp},\vec{P}_{2\perp})$ to $(w,x,\vec{Y},\vec{Z})$.  
Substituting Eq.~(\ref{e:D1q_uDiFF}) or (\ref{e:D1g_uDiFF}) into the r.h.s.~of Eq.~(\ref{e:JuDiFF}) then gives an operator definition of $D_1^{h_1 h_2/i}(w,x,\vec{Y}^2,\vec{Z}^2,\vec{Y}\!\cdot\!\vec{Z})$.
In addition, integrating over one or more of the variables $(w,x,\vec{Y},\vec{Z})$ will define a DiFF that is a number density in the remaining variables.  

For example, if we change variables from $(z_1,z_2,\vec{P}_{1\perp},\vec{P}_{2\perp})$ to $(z,\zeta,\vec{k}_{T},\vec{R}_{T})$, as is typically done when deriving factorization theorems used in phenomenology, then $\mathcal{J}=z^3/2$.  Thus, 
\begin{align}
&D_1^{h_1 h_2/q}(z,\zeta,\vec{k}_T^{\,2},\vec{R}_T^{\,2},\vec{k}_T\cdot \vec{R}_T) \label{e:D1zzetkTRT}\\
&\hspace{0.5cm}= \frac{z}{32\pi^3(1-\zeta^2)}{\rm Tr}\!\left[\Delta^{h_1h_2/q}(z_1,z_2,\vec{P}_{1\perp},\vec{P}_{2\perp})\gamma^-\right]\nonumber
\end{align}
is a number density in $(z,\zeta,\vec{k}_T,\vec{R}_T)$
(where we made use of $z_1z_2=z^2(1-\zeta^2)/4$), and similarly for the gluon case.  Note that the arguments of the correlator on the r.h.s.~can be replaced with $z_{1(2)}=z(1\pm\zeta)/2$ and $\vec{P}_{1(2)\perp} = -z(1\pm\zeta)\vec{k}_T/2\pm\vec{R}_T$.

We emphasize the distinction between our prefactor of $z/(32\pi^3(1-\zeta^2))$ in Eq.~(\ref{e:D1zzetkTRT}) and the prefactor of $1/(4z)$ used by BBJR.  The latter does not allow for the uDiFFs to retain a number density interpretation. The explicit parameterization of Eq.~(\ref{e:DiFF_corr}) in terms of quark and gluon uDiFFs for all parton polarizations, as functions of $(z,\zeta,\vec{k}_T,\vec{R}_T)$, is given in Supplemental Material.

The functions of interest in experimental measurements are the extDiFFs, which we define  by changing variables from $(z_1,z_2,\vec{P}_{1\perp},\vec{P}_{2\perp})$ to $(z,\zeta,\vec{k}_{T},\vec{R}_{T})$ (as above) and integrating over $\vec{k}_T$.  In the quark sector, two twist-2 Dirac projections survive~\cite{Collins:1993kq,Bianconi:1999cd}:
\begin{align}
    &\frac{z}{32\pi^3(1-\zeta^2)}\!\int\!\! d^2\vec{k}_T\,{\rm Tr}\!\left[\Delta^{h_1h_2/q}(z_1,z_2,\vec{P}_{1\perp},\vec{P}_{2\perp})\gamma^-\right]\nonumber\\ 
    &\hspace{2.5cm}= D_1^{h_1h_2/q}(z,\zeta,\vec{R}_T^2)\,,\label{e:extDiFF1}\\
    &\frac{z}{32\pi^3(1-\zeta^2)}\!\int\!\! d^2\vec{k}_T\,{\rm Tr}\!\left[\Delta^{h_1h_2/q}(z_1,z_2,\vec{P}_{1\perp},\vec{P}_{2\perp})i\sigma^{i-}\gamma_5\right] \nonumber\\
    &\hspace{2.5cm}= -\frac{\epsilon_T^{ij}R_T^j}{M_h}H_1^{\sphericalangle\,h_1h_2/q}(z,\zeta,\vec{R}_T^2)\,,\label{e:extDiFF2}
\end{align}
where $\epsilon_T^{ij}=\epsilon^{-+ij}$ with $\epsilon_T^{12}=1$.  One should understand the l.h.s.~of Eqs.~(\ref{e:extDiFF1}), (\ref{e:extDiFF2}) as giving an operator definition of the extDiFFs where the integration over $\vec{k}_T$ has been explicitly carried out on the correlator in Eq.~(\ref{e:DiFF_corr}). In this case, we consider these objects within full QCD.  We note that if one instead changes variables from $(z_1,z_2,\vec{P}_{1\perp},\vec{P}_{2\perp})$ to $(z_1,z_2,\vec{k}_{T},\vec{R}_{T})$, integrating over $\vec{k}_T$ and $\vec{R}_T$ 
leads to the collinear DiFF $D_1^{h_1h_2/q}(z_1,z_2)$, 
and the associated correlator matches that in Refs.~\cite{Majumder:2004br,Majumder:2004wh}. We emphasize the existence of $H_{1}^{\sphericalangle\, h_1h_2/q}(z,\zeta,\vec{R}_T^2)$, which is not present for fragmentation into a single hadron.  This function  has become important in the extraction of the transversity PDFs, which couple to it in dihadron observables~\cite{Bacchetta:2011ip,Courtoy:2012ry,Courtoy:2012ry,Bacchetta:2012ty,Radici:2015mwa, Radici:2016lam,Radici:2018iag, Benel:2019mcq,Cocuzza:2023oam,Cocuzza:2023vqs}. 
The gluon extDiFFs are given in Supplemental Material.

Experimental measurements of dihadron observables are usually differential in $(z,M_h)$ and integrated over~$\zeta$.  
The relevant DiFFs are then dependent on $(z,M_h)$~\cite{Bianconi:1999cd,Radici:2001na,Bacchetta:2002ux,Bacchetta:2003vn,Boer:2003ya,Bacchetta:2004it}.  We change variables from $(z_1,z_2,\vec{P}_{1\perp},\vec{P}_{2\perp})$ to 
$(z,\zeta,\vec{k}_T,M_h,\phi_{R_T})$, where $\phi_{R_T}$ is the azimuthal angle of $\vec{R}_T$.
The Jacobian is $\mathcal{J}=z^3(1-\zeta^2)/8$.  Using our aforementioned prescription, we can define a DiFF that is a number density in $(z,M_h)$: 
\begin{equation}
    D_1^{h_1h_2/i}(z,M_h) \equiv \!\frac{\pi}{2}M_h\!\int_{-1}^1 \!\! d\zeta\,(1-\zeta^2) \! \,D_1^{h_1h_2/i}(z,\zeta,\vec{R}_T^2)\,.\label{e:D1zMh}
\end{equation}
For completeness, we also write down our definition of $H_{1}^{\sphericalangle\, h_1h_2/i}(z,M_h)$: 
\begin{align}
    &H_{1}^{\sphericalangle\, h_1h_2/i}(z,M_h) \label{e:H1zMh}\\
    &\hspace{0.5cm}\equiv \frac{\pi}{2}M_h\!\int_{-1}^1 \!\! d\zeta\,\frac{|\vec{R}_T|}{M_h}\,(1-\zeta^2) \! \,H_{1}^{\sphericalangle\, h_1h_2/i}(z,\zeta,\vec{R}_T^2)\,. \nonumber
\end{align}

Given the number density interpretation of our DiFFs, one can compute expectation values for the ensemble of all $(h_1 h_2)$ pairs in the fragmentation of a parton $i$.  As mentioned, two of the main variables that dihadron measurements are sensitive to are $z$ and $M_h$. The expectation value of an arbitrary function $\mathcal{O}(z,M_h)$ of these variables can then be calculated as
\begin{equation}
    \langle \mathcal{O}(z,M_h)\rangle^{h_1 h_2/i}\!= \!
    \int \!\!dz \,dM_h\,\mathcal{O}(z,M_h)\,D_1^{h_1h_2/i}(z,M_h)\,.\label{e:avg_value_zzetaMh}
\end{equation}
For example, one could compute the average value of $z$ or of $M_h$ for $\pi^+\pi^-$ pairs produced from the fragmentation of an unpolarized quark.
The DiFF $D_1^{h_1h_2/i}(z,M_h)$ can be extracted directly from experiment, e.g., using the cross section  $d\sigma/dz\,dM_h$  for $e^+e^-\to (h_1h_2)X$ measured by Belle~\cite{Belle:2017rwm}.  Our definition of uDiFFs allows us to establish a clear physical meaning for $D_1^{h_1h_2/i}(z,M_h)$, which has been absent thus far in the literature.  

Actually, calculating the leading-order cross section for $d\sigma/dz\, dM_h$ for $e^+e^-\to (h_1h_2)X$ serves as another verification of the number density interpretation of our new definition of $D_1^{h_1h_2/i}(z,M_h)$.  Starting from $P_1^0 P_2^0 \,d\sigma/d^3\vec{P}_1d^3\vec{P}_2$, the result takes the form
\begin{equation}
    \frac{d\sigma}{dz\,dM_h} = \hat{\sigma}^i_0 \, D_1^{h_1h_2/i}(z,M_h)\,.\label{e:e+e-LO}
\end{equation}
For $i=q$, $\hat{\sigma}_0^q=4\pi\alpha_{\rm em}^2N_ce_q^2/(3s)$, which is the partonic cross section for $e^+e^- \to \gamma\to q\bar{q}$, where $\alpha_{\rm em}$ is the fine structure constant, and
$\sqrt{s}$ is the center-of-mass energy of the $e^+e^-$ pair. A sum over quarks and antiquarks is then needed on the r.h.s.~of Eq.~(\ref{e:e+e-LO}).  For $i=g$, $\hat{\sigma}_0^g=[(\alpha_s^2G_F^2)/(576\pi^3)][(m_e^2s^2(N_c^2-1))/(s-m_H^2)^2]$, 
which is the partonic cross section for $e^+e^- \to H\to gg$ ($H$ being the Higgs boson) using an effective $H$-$g$-$g$ coupling~\cite{Ellis:1975ap,Shifman:1979eb,Kniehl:1995tn}, $\alpha_s$ is the strong coupling, $G_F$ is the Fermi constant, and $m_e$ ($m_H$) is the mass of the electron (Higgs).  A factor of~2 is now needed on the r.h.s.~of Eq.~(\ref{e:e+e-LO}) since both gluons have the ability to fragment into the dihadron.  

The structure of Eq.~(\ref{e:e+e-LO}) is exactly what one expects if $D_1^{h_1h_2/i}(z,M_h)$ is to be interpreted as a number density, i.e., the differential cross section equals the partonic cross section times the DiFF.  We have also explicitly confirmed this feature for other sets of variables, including $d\sigma/dz_1dz_2$ and $d\sigma/dz \,d\zeta \,d^2\vec{R}_T$ involving $D_1^{h_1 h_2/i}(z_1,z_2)$ and $D_1^{h_1 h_2/i}(z,\zeta,\vec{R}_T^2)$, respectively.

\begin{figure}
    \centering
    \includegraphics[width=0.475\textwidth]{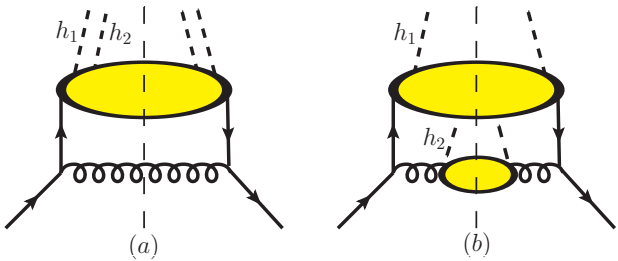}
    \caption{Example diagrams of the (a) homogeneous and (b) inhomogeneous terms for the evolution of the extDiFF $D^{h_1h_2/q}_1(z,\zeta,\vec{R}_T^2)$.}
    \label{f:evo}
\end{figure}

{\it Evolution of Extended DiFFs ---}
Since the extDiFFs in Eqs.~(\ref{e:extDiFF1}), (\ref{e:extDiFF2}) (and their gluon analogues) are the objects that enter most directly in existing phenomenological studies of dihadron observables, it is important to derive their evolution equations for our definition.
Here we analyze the $\mathcal{O}(\alpha_s)$ perturbative corrections to the dihadron fragmentation correlator, similar to what is done  for the single-hadron case -- see, e.g.,  Ref.~\cite{Collins:2011zzd}  Sec.~12.10. The evolution of the DiFF correlator in Eq.~(\ref{e:DiFF_corr}) has two pieces:~a ``homogeneous term'' involving only DiFFs (an example graph is given in Fig.~\ref{f:evo}(a)), and 
an ``inhomogeneous term'' involving single-hadron FFs (an example graph is given in Fig.~\ref{f:evo}(b)).  We have explicitly checked that the inhomogeneous term for the evolution of $D_1^{h_1 h_2/i}(z,\zeta,\vec{R}_T^2)$ is not ultraviolet divergent (see the Supplemental Material), and therefore does not contribute to the evolution of extDiFFs.  The same conclusion was reached in Ref.~\cite{Ceccopieri:2007ip}.  However, this inhomogeneous term is needed to derive the full evolution for the collinear DiFFs $D_1^{h_1 h_2/i}(z_1,z_2)$~\cite{Sukhatme:1980vs,Vendramin:1980wz,Sukhatme:1981ym,deFlorian:2003cg,Majumder:2004br,Majumder:2004wh,Chen:2022pdu,Chen:2022muj}. We also remark that for extDiFFs, inhomogeneous diagrams will contribute at $\mathcal{O}(\alpha_s^2)$ and higher orders of evolution. 

For collinear PDFs and FFs (e.g., $f_1^{i/N}(x)$ and $D_1^{h/i}(z)$), evolution is a perturbative process for  the $1\to 2$ splitting of a parton and is independent of the target (in the case of PDFs) or final state (in the case of FFs) -- see, e.g., Ref.~\cite{Collins:2011zzd} Secs.~9.3.1, 12.9.  This observation, along with the structure of the correlator in Eq.~(\ref{e:DiFF_corr}), the fact that the extDiFFs are obtained by integrating over $\vec{k}_T$, 
and the conclusion that only the homogeneous term contributes to their evolution, makes clear that the splitting functions for extDiFFs will be the same as those for a parton fragmenting into a single hadron. 
The final result reads
\begin{align}
    &\frac{\partial \mathcal{D}^{h_1h_2/i}(z,\zeta,\vec{R}_T^2;\mu)}{\partial \ln\mu^2}\nonumber \\
    &\hspace{0.5cm}= \sum_{i'}\!\int_{z}^1\! \frac{dw}{w}\mathcal{D}^{h_1h_2/i'}\!\!\left(\!\frac{z}{w},\zeta,\vec{R}_T^2;\mu\right)\!P_{i\to i'}(w)\,, \label{e:D1DiFFevo_HOM}
\end{align}
where
$\mathcal{D} = D_1\;{\rm or}\;H_1^\sphericalangle$, and $P_{i\to i'}(w)$ are the unpolarized time-like splitting kernels~\cite{Altarelli:1977zs} when $\mathcal{D} = D_1$, or the transversely polarized splitting kernels~\cite{Stratmann:2001pt} when $\mathcal{D} = H_1^\sphericalangle$.  We note from Eqs.~(\ref{e:D1zMh}), (\ref{e:H1zMh}) it is clear that $D_1^{h_1h_2/i}(z,M_h)$ and $H_{1}^{\sphericalangle\, h_1h_2/i}(z,M_h)$ obey the same evolution equations as Eq.~(\ref{e:D1DiFFevo_HOM}) since the $\zeta$ dependence there is not altered in the evolution.

Evolution equations for extDiFFs were previously derived in Ref.~\cite{Ceccopieri:2007ip} (and commented on in Ref.~\cite{Bacchetta:2008wb}). No correlator was explicitly given, though, preventing us from unambiguously comparing to those results.  We will instead emphasize that the prefactor used to parameterize the dihadron correlator in terms of DiFFs does affect the evolution kernel.  For example, if on the l.h.s.~of Eq.~(\ref{e:D1q_uDiFF}) a prefactor of $1/(4z)$ is used (instead of $1/(64\pi^3z_1z_2)$), then the integrand of Eq.~(\ref{e:D1DiFFevo_HOM}) would not have the $1/w$ factor.
We also mention that the evolution of $D_1^{h_1h_2/i}(z_1,z_2)$ was derived in Refs.~\cite{Sukhatme:1981ym,deFlorian:2003cg,Majumder:2004br,Majumder:2004wh} at leading order and recently in Refs.~\cite{Chen:2022pdu,Chen:2022muj} at next-to-leading order.  We briefly discuss in Supplemental Material how, starting from extDiFFs, one can reproduce the leading-order result.  

{\it Conclusions ---}
We have introduced a new quantum field-theoretic definition for fully unintegrated dihadron fragmentation functions (uDiFFs), as well as a generalized version for $n$-hadron fragmentation, that retains a number density interpretation. We have justified this by proving certain number and momentum sum rules. 
Moreover, we have developed a simple prescription for how to define operators for uDiFFs that are number densities in any variables of interest. In particular, we established a clear physical meaning for the function $D_1^{h_1h_2/i}(z,M_h)$ as a number density in $(z,M_h)$, which was not possible with prior definitions in the literature.  The definitions in Eqs.~(\ref{e:D1q_uDiFF}), (\ref{e:D1g_uDiFF}) will also be beneficial as the starting point for possible factorization theorems beyond leading order for processes involving uDiFFs and extDiFFs.
In addition, we derived the $\mathcal{O}(\alpha_s)$ evolution equations for our extDiFFs. 

With DiFFs now rigorously established as number densities through this work, one can achieve a deeper understanding of hadronic structure through phenomenological extractions of extDiFFs.  Indeed, there is electron-positron annihilation dihadron cross section data~\cite{Belle:2017rwm} available sensitive to $D_1^{h_1h_2/i}(z,M_h)$, as well as several measurements of dihadron transverse single-spin asymmetries  sensitive to $H_1^{\sphericalangle h_1h_2/i}(z,M_h)$ in electron-positron annihilation~\cite{Belle:2011cur}, semi-inclusive deep-inelastic scattering~\cite{HERMES:2008mcr,COMPASS:2023cgk}, and proton-proton collisions~\cite{STAR:2015jkc,STAR:2017wsi}.  A simultaneous global analysis in our framework of all the aforementioned data can be found in Refs.~\cite{Cocuzza:2023oam,Cocuzza:2023vqs}.

\begin{acknowledgments}
{\it Acknowledgments} --- We thank M.~Schlegel for useful discussions regarding the evolution of the DiFFs.
We appreciate fruitful exchanges with T.~C.~Rogers that helped clarify certain aspects of the manuscript.  
This work was supported by the U.S. Department of Energy contract No.~DE-AC05-06OR23177, under which Jefferson Science Associates, LLC operates Jefferson Lab (A.P.~and N.S.), and the National Science Foundation under Grants No.~PHY-2110472 (C.C.~and A.M.), No.~PHY-2011763 and No.~PHY-2308567 (D.P.), and No.~PHY-2012002, No.~PHY-2310031, No.~PHY-2335114 (A.P.).
The work of C.C., A.M., and A.P.~was supported by the U.S. Department of Energy, Office of Science, Office of Nuclear Physics, within the framework of the TMD Topical Collaboration, and by Temple University (C.C. and A.P.).
The work of N.S. was supported by the DOE, Office of Science, Office of Nuclear Physics in the Early Career Program.
\end{acknowledgments}

%

\clearpage
\onecolumngrid
\section{Supplemental Material}
\setcounter{equation}{0}
\renewcommand{\thesubsection}{S\arabic{subsection}}   
\setcounter{secnumdepth}{2}
\renewcommand{\thetable}{S\arabic{table}}   
\renewcommand{\thefigure}{S\arabic{figure}}
\renewcommand{\theequation}{S\arabic{equation}}

\subsection{Parameterization of the uDiFF Correlator}
The twist-2 Dirac traces for $\Delta^{h_1h_2/q}(z_1,z_2,\vec{P}_{1\perp},\vec{P}_{2\perp})$ for unpolarized hadrons, as number densities in $(z_1,z_2,\vec{P}_{1\perp},\vec{P}_{2\perp})$, can be parameterized in terms of the following quark uDiFFs:
\begin{align}
    \frac{1}{64\pi^3z_1 z_2}{\rm Tr}\!\left[\Delta^{h_1h_2/q}(z_1,z_2,\vec{P}_{1\perp},\vec{P}_{2\perp})\gamma^-\right] &= D_1^{h_1h_2/q}(z_1,z_2,\vec{P}_{1\perp}^2,\vec{P}_{2\perp}^2,\vec{P}_{1\perp}\!\cdot \!\vec{P}_{2\perp})\,,\label{e:uDiFF1p}\\[0.3cm]
    \frac{1}{64\pi^3z_1 z_2}{\rm Tr}\!\left[\Delta^{h_1h_2/q}(z_1,z_2,\vec{P}_{1\perp},\vec{P}_{2\perp})\gamma^-\gamma_5\right] &= -\frac{\epsilon_\perp^{ij}R_\perp^i P_{h\perp}^j}{zM_h^2}\,G_1^{\perp\,h_1h_2/q}(z_1,z_2,\vec{P}_{1\perp}^2,\vec{P}_{2\perp}^2,\vec{P}_{1\perp}\!\cdot \!\vec{P}_{2\perp})\,,\label{e:uDiFF2p}\\[0.3cm]
    \frac{1}{64\pi^3z_1 z_2}{\rm Tr}\!\left[\Delta^{h_1h_2/q}(z_1,z_2,\vec{P}_{1\perp},\vec{P}_{2\perp})i\sigma^{i-}\gamma_5\right] &= -\frac{\epsilon_\perp^{ij}R_\perp^j}{M_h}H_1^{\sphericalangle'h_1h_2/q}(z_1,z_2,\vec{P}_{1\perp}^2,\vec{P}_{2\perp}^2,\vec{P}_{1\perp}\!\cdot \!\vec{P}_{2\perp})\nonumber\\
    &\hspace{0.4cm}+\frac{\epsilon_\perp^{ij}P_{h\perp}^j}{zM_h}H_1^{\perp'\,h_1h_2/q}(z_1,z_2,\vec{P}_{1\perp}^2,\vec{P}_{2\perp}^2,\vec{P}_{1\perp}\!\cdot \!\vec{P}_{2\perp})\,,\label{e:uDiFF3p}
\end{align}
where 
$\epsilon_\perp^{ij}=\epsilon^{-+ij}$ with $\epsilon_\perp^{12}=1$, and the momentum components are understood to be in the parton frame.  The correlator $\Delta^{h_1h_2/g}(z_1,z_2,\vec{P}_{1\perp},\vec{P}_{2\perp})$ for unpolarized hadrons, as number densities in $(z_1,z_2,\vec{P}_{1\perp},\vec{P}_{2\perp})$, can be decomposed into the following gluon uDiFFs:
\begin{align}
    \frac{z}{32\pi^3z_1z_2P_h^-}\,\delta_\perp^{ij} \,\Delta^{h_1h_2/g,ij}(z_1,z_2,\vec{P}_{1\perp},\vec{P}_{2\perp}) &= D_1^{h_1h_2/g}(z_1,z_2,\vec{P}_{1\perp}^2,\vec{P}_{2\perp}^2,\vec{P}_{1\perp}\!\cdot \!\vec{P}_{2\perp})\,,\label{e:uDiFFg1p}\\[0.3cm]
    \frac{z}{32\pi^3z_1z_2P_h^-}\,i\epsilon_\perp^{ij} \,\Delta^{h_1h_2/g,ij}(z_1,z_2,\vec{P}_{1\perp},\vec{P}_{2\perp}) &= -\frac{\epsilon_\perp^{ij}R_\perp^i P_{h\perp}^j}{zM_h^2} G_1^{\perp\,h_1h_2/g}(z_1,z_2,\vec{P}_{1\perp}^2,\vec{P}_{2\perp}^2,\vec{P}_{1\perp}\!\cdot \!\vec{P}_{2\perp})\,,\label{e:uDiFFg2p}\\[0.3cm]
    \frac{z}{32\pi^3z_1z_2P_h^-}\,\hat{S}\,\Delta^{h_1h_2/g,ij}(z_1,z_2,\vec{P}_{1\perp},\vec{P}_{2\perp}) &= \hat{S}\left[\frac{R_\perp^iR_\perp^j}{M_h^2} H_1^{\sphericalangle'h_1h_2/g}(z_1,z_2,\vec{P}_{1\perp}^2,\vec{P}_{2\perp}^2,\vec{P}_{1\perp}\!\cdot \!\vec{P}_{2\perp})\right.\nonumber\\
    &\hspace{1cm}+\frac{P_{h\perp}^iP_{h\perp}^j}{z^2M_h^2} H_1^{\perp' h_1h_2/g}(z_1,z_2,\vec{P}_{1\perp}^2,\vec{P}_{2\perp}^2,\vec{P}_{1\perp}\!\cdot \!\vec{P}_{2\perp})\nonumber\\
    &\hspace{1cm}-\left.\frac{R_\perp^iP_{h\perp}^j}{zM_h^2} H_1^{\sphericalangle\perp' h_1h_2/g}(z_1,z_2,\vec{P}_{1\perp}^2,\vec{P}_{2\perp}^2,\vec{P}_{1\perp}\!\cdot \!\vec{P}_{2\perp})\right]\,,\label{e:uDiFFg3p}
\end{align}
where $\hat{S}$ is the symmetrization operator, $\hat{S}O^{ij}\equiv \frac{1}{2}(O^{ij}+O^{ji}-\delta_\perp^{ij}O^{kk})$, in this frame.

The twist-2 Dirac traces for $\Delta^{h_1h_2/q}(z_1,z_2,\vec{P}_{1\perp},\vec{P}_{2\perp})$ for unpolarized hadrons, as number densities in $(z,\zeta,\vec{k}_T,\vec{R}_T)$, can be parameterized in terms of the following quark uDiFFs:
\begin{align}
    \frac{z}{32\pi^3(1-\zeta^2)}{\rm Tr}\!\left[\Delta^{h_1h_2/q}(z_1,z_2,\vec{P}_{1\perp},\vec{P}_{2\perp})\gamma^-\right] &= D_1^{h_1h_2/q}(z,\zeta,\vec{k}_T^{\,2},\vec{R}_T^{\,2},\vec{k}_T\cdot \vec{R}_T)\,,\label{e:uDiFF1h}\\[0.3cm]
    \frac{z}{32\pi^3(1-\zeta^2)}{\rm Tr}\!\left[\Delta^{h_1h_2/q}(z_1,z_2,\vec{P}_{1\perp},\vec{P}_{2\perp})\gamma^-\gamma_5\right] &= \frac{\epsilon_T^{ij}R_T^ik_T^j}{M_h^2}\,G_1^{\perp\,h_1h_2/q}(z,\zeta,\vec{k}_T^{\,2},\vec{R}_T^{\,2},\vec{k}_T\cdot \vec{R}_T)\,,\label{e:uDiFF2h}\\[0.3cm]
    \frac{z}{32\pi^3(1-\zeta^2)}{\rm Tr}\!\left[\Delta^{h_1h_2/q}(z_1,z_2,\vec{P}_{1\perp},\vec{P}_{2\perp})i\sigma^{i-}\gamma_5\right] &= -\frac{\epsilon_T^{ij}R_T^j}{M_h}H_1^{\sphericalangle'h_1h_2/q}(z,\zeta,\vec{k}_T^{\,2},\vec{R}_T^{\,2},\vec{k}_T\cdot \vec{R}_T)\nonumber\\
    &\hspace{0.4cm}-\frac{\epsilon_T^{ij}k_T^j}{M_h}H_1^{\perp'\,h_1h_2/q}(z,\zeta,\vec{k}_T^{\,2},\vec{R}_T^{\,2},\vec{k}_T\cdot \vec{R}_T)\,,\label{e:uDiFF3h}
\end{align}  
where $\epsilon_T^{ij}=\epsilon^{-+ij}$ with $\epsilon_T^{12}=1$, and the momentum components are understood to be in the dihadron frame. 
The correlator $\Delta^{h_1h_2/g}(z_1,z_2,\vec{P}_{1\perp},\vec{P}_{2\perp})$ for unpolarized hadrons, as number densities in $(z,\zeta,\vec{k}_T,\vec{R}_T)$, can be decomposed into the following gluon uDiFFs:
\begin{align}
    \frac{z^2}{16\pi^3(1-\zeta^2)P_h^-}\,\delta_T^{ij} \,\Delta^{h_1h_2/g,ij}(z_1,z_2,\vec{P}_{1\perp},\vec{P}_{2\perp}) &= D_1^{h_1h_2/g}(z,\zeta,\vec{k}_T^{\,2},\vec{R}_T^{\,2},\vec{k}_T\cdot \vec{R}_T)\,,\label{e:uDiFFg1h}\\[0.3cm]
    \frac{z^2}{16\pi^3(1-\zeta^2)P_h^-}\,i\epsilon_T^{ij} \,\Delta^{h_1h_2/g,ij}(z_1,z_2,\vec{P}_{1\perp},\vec{P}_{2\perp}) &= \frac{\epsilon_T^{ij}R_T^ik_T^j}{M_h^2} G_1^{\perp\,h_1h_2/g}(z,\zeta,\vec{k}_T^{\,2},\vec{R}_T^{\,2},\vec{k}_T\cdot \vec{R}_T)\,,\label{e:uDiFFg2h}\\[0.3cm]
    \frac{z^2}{16\pi^3(1-\zeta^2)P_h^-}\,\hat{S}\,\Delta^{h_1h_2/g,ij}(z_1,z_2,\vec{P}_{1\perp},\vec{P}_{2\perp}) &= \hat{S}\left[\frac{R_T^iR_T^j}{M_h^2} H_1^{\sphericalangle'h_1h_2/g}(z,\zeta,\vec{k}_T^{\,2},\vec{R}_T^{\,2},\vec{k}_T\cdot \vec{R}_T)\right.\nonumber\\
    &\hspace{1cm}+\frac{k_T^ik_T^j}{M_h^2} H_1^{\perp' h_1h_2/g}(z,\zeta,\vec{k}_T^{\,2},\vec{R}_T^{\,2},\vec{k}_T\cdot \vec{R}_T)\nonumber\\
    &\hspace{1cm}+\left.\frac{R_T^ik_T^j}{M_h^2} H_1^{\sphericalangle\perp' h_1h_2/g}(z,\zeta,\vec{k}_T^{\,2},\vec{R}_T^{\,2},\vec{k}_T\cdot \vec{R}_T)\right]\,,\label{e:uDiFFg3h}
\end{align}
where $\hat{S}O^{ij}\equiv \frac{1}{2}(O^{ij}+O^{ji}-\delta_T^{ij}O^{kk})$ in this frame.  We note that the structures on the r.h.s.~of Eqs.~(\ref{e:uDiFF1h})--(\ref{e:uDiFFg2h}) were first written down in Ref.~\cite{Bianconi:1999cd} while the r.h.s.~of Eq.~(\ref{e:uDiFFg3h}) first appeared in Ref.~\cite{Metz:2016swz}.

The quark extDiFFs were already provided in the main text but are given here again for completeness:
\begin{align}
    &\frac{z}{32\pi^3(1-\zeta^2)}\!\int\!\! d^2\vec{k}_T\,{\rm Tr}\!\left[\Delta^{h_1h_2/q}(z_1,z_2,\vec{P}_{1\perp},\vec{P}_{2\perp})\gamma^-\right]= D_1^{h_1h_2/q}(z,\zeta,\vec{R}_T^2)\,,\label{e:extDiFF1_SM}\\
    &\frac{z}{32\pi^3(1-\zeta^2)}\!\int\!\! d^2\vec{k}_T\,{\rm Tr}\!\left[\Delta^{h_1h_2/q}(z_1,z_2,\vec{P}_{1\perp},\vec{P}_{2\perp})i\sigma^{i-}\gamma_5\right] = -\frac{\epsilon_T^{ij}R_T^j}{M_h}H_1^{\sphericalangle\,h_1h_2/q}(z,\zeta,\vec{R}_T^2)\,. \label{e:extDiFF2_SM}
\end{align}
The gluon extDiFFs are given by
\begin{align}
    &\frac{z^2}{16\pi^3(1-\zeta^2)P_h^-}\delta_T^{ij}\!\!\int \!\!d^2\vec{k}_T\,\Delta^{h_1h_2/g,ij}(z_1,z_2,\vec{P}_{1\perp},\vec{P}_{2\perp}) = D_1^{h_1h_2/g}(z,\zeta,\vec{R}_T^2)\,,\label{e:extDiFF3_SM}\\[0.3cm]
    &\frac{z^2}{16\pi^3(1-\zeta^2)P_h^-}\,\hat{S}\!\!\int \!\!d^2\vec{k}_T\,\Delta^{h_1h_2/g,ij}(z_1,z_2,\vec{P}_{1\perp},\vec{P}_{2\perp}) = \hat{S}\left[\frac{R_T^iR_T^j}{M_h^2} H_1^{\sphericalangle h_1h_2/g}(z,\zeta,\vec{R}_T^2)\right]. \label{e:extDiFF4_SM}
\end{align}
One should understand the l.h.s.~of Eqs.~(\ref{e:extDiFF1_SM})--(\ref{e:extDiFF4_SM}) as giving an operator definition of the extDiFFs where the integration over $\vec{k}_T$ has been explicitly carried out on the correlator in Eq.~(\ref{e:DiFF_corr}).  That is, on the l.h.s.~of Eqs.~(\ref{e:extDiFF1_SM}), (\ref{e:extDiFF2_SM}),
\begin{align}
\int\!&d^2\vec{k}_T\,\Delta^{h_1h_2/q}_{\alpha\beta}(z_1,z_2,\vec{P}_{1\perp},\vec{P}_{2\perp})\nonumber\\
&\equiv \frac{1}{N_i}\!\sum_X\hspace{-0.5cm}\int\, \int\!\frac{d\xi^+}{2\pi}\,e^{ik^-\xi^+}\langle 0|\mathcal{W}(\infty^+,\xi^+)\psi_{q,\alpha}(\xi^+,0^-,\vec{0}_\perp)|P_1,P_2;X\rangle \langle P_1,P_2;X|\bar{\psi}_{q,\beta}(0^+,0^-,\vec{0}_\perp)\mathcal{W}(0^+,\infty^+)|0\rangle\,,
\end{align}
and similarly for Eqs.~(\ref{e:extDiFF3_SM}), (\ref{e:extDiFF4_SM}).

We note that if one instead changes variables from $(z_1,z_2,\vec{P}_{1\perp},\vec{P}_{2\perp})$ to $(z_1,z_2,\vec{k}_{T},\vec{R}_{T})$, integrating over $\vec{k}_T$ and $\vec{R}_T$ 
leads to the collinear DiFFs $D_1^{h_1h_2/i}(z_1,z_2)$, 
and the associated correlators match those in Refs.~\cite{Majumder:2004br,Majumder:2004wh}.

\subsection{Proof of Number Sum Rule for  $\mathbf{D_1^{h_1h_2/i}}$}
We start by first proving the number sum rule for the single-hadron TMD FF, focusing on the  quark case $D_1^{h/q}(z,\vec{P}_\perp^2)$.   The essential elements are similar to the derivation of the single-hadron momentum sum rule carried out in Ref.~\cite{Meissner:2010cc}. The correlator definition for this function in the parton frame is given by
\begin{equation}
    D_1^{h/q}(z,\vec{P}_\perp^2)= \frac{1}{N_c}\frac{1}{4z}\sum_X\hspace{-0.5cm}\int\,\int\! \frac{d\xi^+d^2\vec{\xi}_\perp}{(2\pi)^3}\,e^{ik^-\xi^+}{\rm Tr}\bigg[\!\langle 0|\mathcal{W}(\infty,\xi)\psi_{q}(\xi^+,0^-,\vec{\xi}_\perp)|P;X\rangle\langle P;X|\bar{\psi}_{q}(0^+,0^-,\vec{0}_\perp)\mathcal{W}(0,\infty)|0\rangle\,\gamma^-\!\bigg],
\end{equation}
where a sum over colors is implied.  We then have
\begin{equation}
    \sum_h\!\int_0^1 \!\!dz\!\int\! \!d^2\!\vec{P}_\perp D_1^{h/q}(z,\vec{P}_\perp^2)= \frac{1}{N_c}\frac{1}{2}\int\!\! d\xi^+d^2\vec{\xi}_\perp\,e^{ik^-\xi^+}{\rm Tr}\bigg[\!\langle 0|\mathcal{W}(\infty,\xi)\psi_{q}(\xi^+,0^-,\vec{\xi}_\perp)\hat{N}\bar{\psi}_{q}(0^+,0^-,\vec{0}_\perp)\mathcal{W}(0,\infty)|0\rangle\,\gamma^-\!\bigg],
\end{equation}
where we have made use of the completeness relation ${\scriptstyle\sum_X}\hspace{-0.55cm}\int \;\;\,|X\rangle\langle X|=1$ and denoted the total number operator by
\begin{equation}
    \hat{N} \equiv \sum_h\int\!\!\frac{dP^-d^2\vec{P}_{\perp}}{(2\pi)^3 \,2P^-}\,\hat{a}_{h}^\dagger \hat{a}_{h}\,.\label{e:numop}
\end{equation}
Using the so-called ``good'' quark field $\psi_{-,q} \equiv \frac{1}{2}\gamma^+\gamma^-\psi_q$ leads to
\begin{align}
    \sum_h\int_0^1 \!dz\! &\int \!d^2\vec{P}_\perp \,D_1^{h/q}(z,\vec{P}_\perp^2) \nonumber \\
    &= \langle\mathcal{N}\rangle\,\frac{1}{N_c}\frac{1}{\sqrt{2}}\int\! \!d\xi^+d^2\vec{\xi}_\perp\,e^{ik^-\xi^+}{\rm Tr}\bigg[\langle 0|\mathcal{W}(\infty,\xi)\{\psi_{-,q}(\xi^+,0^-,\vec{\xi}_\perp),\psi^\dagger_{-,q}(0^+,0^-,\vec{0}_\perp)\}\mathcal{W}(0,\infty)|0\rangle\,\gamma^-\bigg],\label{e:step1}
\end{align}
where we are able to insert the anticommutator of the good quark fields~\cite{Meissner:2010cc}. The quantity $\langle\mathcal{N}\rangle$ is the expectation value for the total number of hadrons produced by the fragmentation of the quark.  Since
\begin{equation}
    \{\psi_{-,q}(\xi^+,0^-,\vec{\xi}_\perp),\psi^\dagger_{-,q}(0^+,0^-,\vec{0}_\perp)\} = \frac{1}{2\sqrt{2}}\gamma^+\gamma^-\delta(\xi^+)\delta^{(2)}(\vec{\xi}_\perp)\,, \label{e:step2}
\end{equation}
we arrive at the result (see also Ref.~\cite{Ito:2009zc})
\begin{equation}
    \sum_h\int_0^1 \!dz\!\int \!\!d^2\vec{P}_\perp \,D_1^{h/q}(z,\vec{P}_\perp^2)= \langle\mathcal{N}\rangle\,.\label{e:num_single}
\end{equation}
An analogous sum rule holds for the gluon TMD FF.

We next prove the number sum rule for the uDiFF  $D_1^{h_1h_2/q}(z_1,z_2,\vec{P}_{1\perp}^2,\vec{P}_{2\perp}^2,\vec{P}_{1\perp}\!\cdot \!\vec{P}_{2\perp})$. 
We start with
\begin{align}
    &\sum_{h_1}\sum_{h_2}\int  \!dz_1\,d^2\vec{P}_{1\perp}\!\int\! \!dz_2\,d^2\vec{P}_{2\perp} \frac{1}{64\pi^3z_1 z_2} {\rm Tr}\bigg[\Delta^{h_1h_2/q}(z_1,z_2,\vec{P}_{1\perp},\vec{P}_{2\perp})\gamma^-\bigg]
    \nonumber\\
    &= {\sum_{h_1}}\hspace{-0.45cm}\int{\sum_{h_2}}\hspace{-0.45cm}\int \,\frac{1}{N_c}\frac{1}{2}\int\! \!d\xi^+d^2\vec{\xi}_\perp\,e^{ik^-\xi^+} {\rm Tr}\bigg[\langle 0|\mathcal{W}(\infty,\xi)\psi_{q}(\xi^+,0^-,\vec{\xi}_\perp)\hat{a}_{h_2}^\dagger \hat{a}_{h_1}^\dagger \hat{a}_{h_1}\hat{a}_{h_2}\bar{\psi}_{q}(0^+,0^-,\vec{0}_\perp)\mathcal{W}(0,\infty)|0\rangle\,\gamma^-\bigg],\label{e:D1sumrule1}
\end{align}
where both sums run over all hadron types for $h_1$ and $h_2$, and we have made use of the completeness relation ${\scriptstyle\sum_X}\hspace{-0.55cm}\int \;\;\,|X\rangle\langle X|=1$.  Using the commutation relations between creation and annihilation operators, and inserting the uDiFF on the l.h.s., we eventually arrive at
\begin{align}
    &\sum_{h_1}\sum_{h_2}\int  \!dz_1\,d^2\vec{P}_{1\perp}\!\int\! \!dz_2\,d^2\vec{P}_{2\perp} D_1^{h_1h_2/q}(z_1,z_2,\vec{P}_{1\perp}^2,\vec{P}_{2\perp}^2,\vec{P}_{1\perp}\cdot \vec{P}_{2\perp})
    \nonumber\\
    &= \frac{1}{N_c}\frac{1}{2}\int\! \!d\xi^+d^2\vec{\xi}_\perp\,e^{ik^-\xi^+} {\rm Tr}\bigg[\langle 0|\mathcal{W}(\infty,\xi)\psi_{q}(\xi^+,0^-,\vec{\xi}_\perp)\bigg(\!\sum_{h_1}\sum_{h_2}\hat{N}_{h_1}\hat{N}_{h_2}-\sum_{h_1}\hat{N}_{h_1}\!\bigg)\bar{\psi}_{q}(0^+,0^-,\vec{0}_\perp)\mathcal{W}(0,\infty)|0\rangle\,\gamma^-\bigg],\label{e:D1sumrule2}
\end{align}
where we have introduced the number operator, 
\begin{equation}
    \hat{N}_{h_j} \equiv \int\!\frac{dP_j^-d^2\vec{P}_{j\perp}}{(2\pi)^3 \,2P_j^-}\,\hat{a}_{h_j}^\dagger \hat{a}_{h_j}\,,
\end{equation}
for each hadron ($j=1\;{\rm or}\;2$).  Using the definition of the total number operator (\ref{e:numop}),
we find 
\begin{align}
    \sum_{h_1}\sum_{h_2}&\int  \!dz_1\,d^2\vec{P}_{1\perp}\!\int\! \!dz_2\,d^2\vec{P}_{2\perp} D_1^{h_1h_2/q}(z_1,z_2,\vec{P}_{1\perp}^2,\vec{P}_{2\perp}^2,\vec{P}_{1\perp}\cdot \vec{P}_{2\perp})
    \nonumber\\
    &= \frac{1}{N_c}\frac{1}{2}\int\! \!d\xi^+d^2\vec{\xi}_\perp\,e^{ik^-\xi^+} {\rm Tr}\bigg[\langle 0|\mathcal{W}(\infty,\xi)\psi_{q}(\xi^+,0^-,\vec{\xi}_\perp)\hat{N}(\hat{N}-1)\bar{\psi}_{q}(0^+,0^-,\vec{0}_\perp)\mathcal{W}(0,\infty)|0\rangle\,\gamma^-\bigg].\label{e:D1sumrule2}
\end{align}
Following the same steps (\ref{e:step1}), (\ref{e:step2}) as the single-hadron case, we finally arrive at (see also Ref.~\cite{Majumder:2004br})
\begin{equation}
   \sum_{h_1}\sum_{h_2}\int_0^1 dz_2\int_0^{1-z_2}\!dz_1\!\!\int\! \!d^2\vec{P}_{1\perp}\!\int\! \!d^2\vec{P}_{2\perp} D_1^{h_1h_2/q}(z_1,z_2,\vec{P}_{1\perp}^2,\vec{P}_{2\perp}^2,\vec{P}_{1\perp}\cdot \vec{P}_{2\perp})
    = \langle \mathcal{N}(\mathcal{N}-1)\rangle\,,
\end{equation}  
where the support properties restrict the upper limit of the integral on the l.h.s.~to be $(1-z_2)$. The quantity $\langle\mathcal{N}(\mathcal{N}-1)\rangle$ is the expectation value for the total number of hadron pairs produced by the fragmentation of the quark.  An analogous sum rule holds for the gluon uDiFF.

For the $n$-hadron fragmentation case the corresponding number sum rule reads
\begin{equation}
   \sum_{h_1,\dots, h_n}\!\int\!\! dz_n\cdots dz_1 \!\!\int\! \!d^2\vec{P}_{1\perp}\cdots d^2\vec{P}_{n\perp}\, D_1^{\{h_i\}_n/i}(\{z_i\}_n,\{\vec{P}_{i\perp}^2\}_n,\{\vec{P}_{i\perp}\!\cdot\vec{P}_{j\perp}\}_n)
    = \left\langle\prod_{k=0}^{n-1}(\mathcal{N}-k)\!\right\rangle \,,
    \label{e:num_sumrule_n}
\end{equation}  
which can be derived by showing that 
\begin{equation}
{\sum_{h_1}}\hspace{-0.45cm}\int\cdots {\sum_{h_{n}}}\hspace{-0.45cm}\int\hat{a}^\dagger_{h_n}\hat{a}^\dagger_{h_{n-1}}\!\cdots\hat{a}^\dagger_{h_1}\hat{a}_{h_1}\hat{a}_{h_2}\cdots \hat{a}_{h_n} = \prod_{k=0}^{n-1}(\hat{N}-k)\,,\label{e:ncaop}
\end{equation}
and then following the same steps (\ref{e:step1}), (\ref{e:step2}) as the single-hadron case.  We will proceed through a proof by induction.  We have already shown the base cases $n=1$ and $n=2$ to be true.  Let us assume that Eq.~(\ref{e:ncaop}) holds up to $n$ hadrons.  Consider the $(n+1)$-hadron case:
\begin{equation}
{\sum_{h_1}}\hspace{-0.45cm}\int\cdots {\sum_{h_{n+1}}}\hspace{-0.55cm}\int\hat{a}^\dagger_{h_{n+1}}\hat{a}^\dagger_{h_{n}}\cdots\hat{a}^\dagger_{h_1}\hat{a}_{h_1}\hat{a}_{h_2}\cdots \hat{a}_{h_{n+1}} = {\sum_{h_{n+1}}}\hspace{-0.55cm}\int\hat{a}^\dagger_{h_{n+1}}\prod_{k=0}^{n-1}(\hat{N}-k)\hat{a}_{h_{n+1}}\,,\label{e:np1caop}
\end{equation} 
where we have invoked the induction hypothesis to obtain the r.h.s.  Consider this operator acting on an eigenstate $|\mathcal{N}\rangle$ (with eigenvalue $\mathcal{N}$) of the total number operator $\hat{N}$:
\begin{equation}
{\sum_{h_{n+1}}}\hspace{-0.55cm}\int\hat{a}^\dagger_{h_{n+1}}\prod_{k=0}^{n-1}(\hat{N}-k)\hat{a}_{h_{n+1}}|\mathcal{N}\rangle = {\sum_{h_{n+1}}}\hspace{-0.55cm}\int\hat{a}^\dagger_{h_{n+1}}\hat{a}_{h_{n+1}}|\mathcal{N}\rangle\prod_{k=0}^{n-1}((\mathcal{N}-1)-k)=\prod_{k=0}^{n}(\mathcal{N}-k)|\mathcal{N}\rangle = \prod_{k=0}^{n}(\hat{N}-k)|\mathcal{N}\rangle\,, \label{e:np1caop2}
\end{equation}
where for the first equality we have used that $\hat{N}(\hat{a}_{h}|\mathcal{N}\rangle)= (\mathcal{N}-1)(\hat{a}_{h}|\mathcal{N}\rangle)$ for any hadron type $h$, and for the second equality we again utilized the definition of the total number operator (\ref{e:numop}).  Since an operator is entirely determined by its action on a complete set of eigenstates, Eq.~(\ref{e:np1caop2}) establishes Eq.~(\ref{e:ncaop}) holds for $(n+1)$ hadrons, which completes the proof.

\subsection{Proof of Momentum Sum Rule for $\mathbf{D_1^{h_1h_2/i}}$}
We now prove a momentum sum rule for the uDiFF $D_1^{h_1h_2/i}(z_1,z_2,\vec{P}_{1\perp}^2,\vec{P}_{2\perp}^2,\vec{P}_{1\perp}\cdot \vec{P}_{2\perp})$.
We follow steps similar to the derivation of the single-hadron momentum sum rule carried out in Ref.~\cite{Meissner:2010cc}.  
Working in the parton frame and focusing on the quark case, we write 
\begin{align}
    \sum_{h_1}&\int_0^1\!dz_1\!\!\int\! \!d^2\vec{P}_{1\perp} P_1^\mu\, \frac{1}{64\pi^3z_1 z_2} {\rm Tr}\!\bigg[\Delta^{h_1h_2/q}(z_1,z_2,\vec{P}_{1\perp},\vec{P}_{2\perp})\gamma^-\bigg]
    \nonumber\\
    &= \sum_{h_1}\!\frac{1}{N_c\,32\pi^3 z_2}\int\! \!d\xi^+d^2\vec{\xi}_\perp\,e^{ik^-\xi^+} {\rm Tr}\!\bigg[\langle 0|\mathcal{W}(\infty,\xi)\psi_{q}(\xi^+,0^-,\vec{\xi}_\perp)\hat{a}_{h_2}^\dagger\hat{a}_{h_1}^\dagger P_{1}^\mu \hat{a}_{h_1}\hat{a}_{h_2}\bar{\psi}_{q}(0^+,0^-,\vec{0}_\perp)\mathcal{W}(0,\infty)|0\rangle\,\gamma^-\bigg],\label{e:D1sumrule1}
\end{align}
where 
the sum runs over all hadron types for $h_1$, and we have made use of the completeness relation ${\scriptstyle\sum_X}\hspace{-0.55cm}\int \;\;\,|X\rangle\langle X|=1$. 
Again using the commutation relations between creation and annihilation operators, we eventually arrive at
\begin{align}
    \sum_{h_1}\int_0^1\!dz_1\!\!\int\! \!d^2\vec{P}_{1\perp} P_1^\mu\, &\frac{1}{64\pi^3z_1 z_2} {\rm Tr}\!\bigg[\Delta^{h_1h_2/q}(z_1,z_2,\vec{P}_{1\perp},\vec{P}_{2\perp})\gamma^-\bigg]
    \nonumber\\
    &= \frac{1}{N_c\,32\pi^3 z_2}\sum_X\hspace{-0.5cm}\int\,\int \!\!d\xi^+d^2\vec{\xi}_\perp\,e^{ik^-\xi^+}{\rm Tr}\!\bigg[\langle 0|\mathcal{W}(\infty,\xi)\psi_{q}(\xi^+,0^-,\vec{\xi}_\perp)(\hat{P}^\mu\!-\!P_{2}^\mu) \!|P_2;X\rangle\nonumber \\[-0.3cm]
    &\hspace{6.45cm} \times\,\langle P_2;X|\bar{\psi}_{q}(0^+,0^-,\vec{0}_\perp)\mathcal{W}(0,\infty)|0\rangle\,\gamma^-\bigg],\label{e:D1sum_step}
\end{align}
where we have introduced the total momentum operator
\begin{equation}
    \hat{P}^\mu \equiv \sum_h\int\!\frac{dP^-d^2\vec{P}_{\perp}}{(2\pi)^3 \,2P^-}\,\hat{a}_{h}^\dagger P^\mu \hat{a}_{h}\,.
\end{equation}
Using the identity
$\langle 0|\mathcal{W}(\infty,\xi)\psi_{q}(\xi)\hat{P}^\mu$ = $i\partial^\mu\big[ \!\langle 0|\mathcal{W}(\infty,\xi)\psi_{q}(\xi)\!\big]$,
integrating by parts, and choosing $\mu= -$ gives
\begin{equation}
   \sum_{h_1}\!\! \int_0^{1-z_2} \!\!\!dz_1 \!\!\int \!\!d^2\vec{P}_{1\perp}\,z_1\,D_1^{h_1h_2/q}(z_1,z_2,\vec{P}_{1\perp}^{\,2},\vec{P}_{2\perp}^{\,2},\vec{P}_{1\perp}\!\!\cdot\! \vec{P}_{2\perp})= (1-z_2)\,D_1^{h_2/q}(z_2,\vec{P}_{2\perp}^{\,2})\,,\label{e:sumrule_unintS}
\end{equation}
where we have now inserted the uDiFF on the l.h.s.~and single-hadron TMD FF~\cite{Collins:1981uw,Mulders:1995dh,Boer:2003cm,Bacchetta:2006tn} on the r.h.s.  The support properties restrict the upper limit of the integral on the l.h.s.~to be $(1-z_2)$.  
An analogous sum rule holds involving the gluon uDiFF and gluon TMD FF.  In addition, if either or both hadrons have nonzero spin, then a sum over the spin of $h_1$ must be included on the l.h.s.

\subsection{Evolution of Collinear DiFFs}
We briefly discuss how to reproduce the evolution of $D_1^{h_1 h_2/i}(z_1,z_2)$ starting from the evolution of extended DiFFs. The homogeneous part of the evolution for $D_1^{h_1 h_2/i}(z_1,z_2,\vec{R}_T^2)$ has the same structure as Eq.~(\ref{e:D1DiFFevo_HOM}) but with a $1/w^2$ in the integrand on the r.h.s.~instead of $1/w$. One then integrates over $\vec{R}_T$, which trivially becomes an evolution involving $D_1^{h_1h_2/q}(z_1,z_2)$. The complete evolution of $D_1^{h_1h_2/i}(z_1,z_2)$ also requires the inhomogeneous diagrams (integrated over $\vec{R}_T$).  For example, at leading twist the graph in Fig.~\ref{f:evo}(b) before integration gives
\begin{align}
    &D^{h_1h_2/q,{\rm Fig.\!\!~\ref{f:evo}(b)}}_1(z_1,z_2,\vec{R}_T^2;\mu)=\frac{1}{\vec{R}_T^2}\frac{C_F\alpha_s}{2\pi^2}\mu^{2\epsilon}\label{e:D1DiFFevo_INHOM}\int_{z_1}^{1-z_2}\!\!\!\!\frac{dw}{w(1-w)}D_1^{h_1/q}(z_1/w)D_1^{h_2/g}(z_2/(1-w))\frac{1+w^2}{1-w}\,,
\end{align}
where $C_F=(N_c^2-1)/(2N_c)$ and $\alpha_s=g^2/(4\pi)$.  We work in $4-2\epsilon$ dimensions, where the strong coupling is $g\mu^\epsilon$, and for brevity do not include terms of $\mathcal{O}(\epsilon)$.  Note that Eq.~(\ref{e:D1DiFFevo_INHOM}) has no $1/\epsilon$ divergence, explicitly demonstrating that the inhomogeneous term does not contribute to the evolution of extDiFFs. Integrating over $\vec{R}_T$ in dimensional regularization eventually leads to the following as $\epsilon \to 0$:
\begin{align}
    &\frac{\partial D^{h_1h_2/q,{\rm Fig.\!\!~\ref{f:evo}(b)}}_1(z_1,z_2;\mu)}{\partial \ln\mu^2}=\frac{C_F\alpha_s}{2\pi}\int_{z_1}^{1-z_2}\!\!\!\!\frac{dw}{w(1-w)}D_1^{h_1/q}(z_1/w)D_1^{h_2/g}(z_2/(1-w))\frac{1+w^2}{1-w}\,.
\end{align}
The factor $(1+w^2)/(1-w)$ exactly agrees with the real contribution to $q\to qg$ splitting for the evolution of $D_1^{h/q}(z)$.  This pattern generalizes to the other inhomogeneous terms in the evolution for the quark and gluon collinear DiFFs $D_1^{h_1h_2/i}(z_1,z_2)$~\cite{Sukhatme:1981ym,deFlorian:2003cg,Majumder:2004br,Majumder:2004wh,Chen:2022pdu,Chen:2022muj}.

\end{document}